\documentclass{phb-proc4-auth}

\usepackage{graphics}
\usepackage{graphicx}
\usepackage{amssymb}

\begin{document}
\begin{frontmatter}

\journal{SCES '04, version 1}

\title{Unconventional superconductivity and magnetism
in  $\rm CePt_3Si_{1-x}Ge_x$}

\author[TU]{E. Bauer\corauthref{Bau},}
\ead{bauer@ifp.tuwien.ac.at}
\author[TU]{G. Hilscher,}
\author[TU]{H. Michor,}
\author[TU]{M. Sieberer,}
\author[A]{E.W. Scheidt,}
\author[MO]{A. Gribanov,}
\author[MO]{Yu. Seropegin,}
\author[Uni]{P. Rogl,}
\author[PSI]{A. Amato,}
\author[ISIS,Korea]{W.Y.Song,}
\author[Korea]{J.-G. Park,}
\author[ISIS]{D.T. Adroja,}
\author[Dresden]{M. Nicklas,}
\author[Dresden]{G. Sparn,}
\author[Osaka]{M. Yogi,}
\author[Osaka]{and Y. Kitaoka}

\address[TU]{Institut f\"ur Festk\"orperphysik,
Technische Universit\"at Wien, A-1040 Wien, Austria}

\address[A]{Chemische Physik und Materialwissenschaften,
Universit\"at Augsburg, D - 86159 Augsburg, Germany}

\address[MO]{Department of Chemistry, Moscow State University,
Moscow, Russia}

\address[Uni]{Institut f\"ur Physikalische Chemie, Universit\"at Wien,
            A-1090 Wien, Austria}
			
\address[PSI]{Laboratory for Muon-Spin Spectroscopy, 
Paul Scherrer Institute, CH-5232 Villigen PSI, Switzerland }
			
\address[ISIS]{
ISIS Facility, Rutherford Appleton Laboratory, Oxon OX11 0QX,
England}

\address[Korea]{
Department of Physics and Institute of Basic Science,
Sungkyunkwan University, Suwon 440-746, Korea}

\address[Dresden]{Max Planck Institute for Chemical Physics of Solids,
D-011187 Dresden, Germany}

\address[Osaka]{Department of Materials Science and Technology,
Osaka University, Osaka 560-8531, Japan}

\corauth[Bau]{Corresponding author. Tel: +43 1 58 801 131 60
fax: +43 1 58 801 131 99}

\begin{abstract}

$\rm CePt_3Si$ is a novel ternary compound
 exhibiting antiferromagnetic order at $T_N \approx 2.2$~K
and superconductivity (SC) at $T_c \approx 0.75$~K. Large values
of $H_{c2}' \approx -8.5$~T/K and $H_{c2}(0) \approx 5$~T
indicate Cooper pairs formed out of heavy quasiparticles. The
mass enhancement originates from Kondo interaction with a
characteristic temperature $T_K \approx 8$~K. NMR and $\mu$SR
measurements evidence coexistence of SC and long range
magnetic order on a microscopic scale.
Moreover, $\rm CePt_3Si$  is the
first heavy fermion SC without an inversion symmetry. This gives
rise to a novel type of the NMR relaxation rate $1/T_1$ which is
very unique and never reported before for other heavy fermion
superconductors. Studies of Si/Ge substitution allow us to
establish a phase diagram.

\end{abstract}


\end{frontmatter}


Strongly correlated electron systems have been one of the most
interesting topics in condensed matter physics. The key importance
in such research is undoubtedly the discovery of unexpected
features and new phases of metals, intermetallics and oxides at low
temperatures. Among the topics, 
quantum phase transitions and related quantum critical
phenomena are of particular importance. Quantum critical
fluctuations can lead to strong renormalization of normal metallic
properties as well as to novel exotic phases emerging from these
strongly fluctuating environments. One of the most exciting
features in this context is the occurrence of SC.

A recently discovered example is tetragonal
$\rm CePt_3Si$ \cite{Bauer}, the first heavy fermion 
SC without a centre  of inversion.  This implies that 
the electron bands are non-degenerate, except 
along some high-symmetry lines in the Brillouin zone
\cite{Samo}.  
Reduced degeneracy, however,
 weakens or even suppresses SC. Superconductivity
with spin-triplet pairing 
should require inversion symmetry to obtain the necessary
degenerate electron states \cite{Anderson}. 
Thus, it became a
widespread view that a material lacking an inversion
center would be an unlikely candidate for spin-triplet
pairing \cite{Sigrist}. Nevertheless, 
the extreme large value of the upper critical field
$H_{c2}(0) \approx 5$~T of $\rm CePt_3Si$ 
is inconsistent with  
spin-singlet Cooper-pairs, hinting at some novel features
of the SC order parameter.  

The aim of this paper is to provide a review of current research on
$\rm CePt_3Si$
and to locate the system  in the standard
generic phase diagram of heavy fermion compounds through 
both substitution and pressure experiments.
The paper is organised as follows: After a discussion of  normal
state properties of $\rm CePt_3Si$, the SC features
of $\rm CePt_3Si$ are examined before reporting the evolution of
magnetism and SC in $\rm CePt_3(Si,Ge)$.

\subsection*{Normal state properties}

Physical properties of ternary $\rm CePt_3Si$ are dominated by 
long range magnetic order below $T_N \approx 2.2$~K and  
SC below $T_c = 0.75$~K. 
Crystal electric field splitting (CEF) and Kondo interaction 
substantially modify Hund's $J=5/2$ ground state of the Ce ion.
The response of the system associated with 
the mutual interplay of these phenomena will
be highlighted below.

\begin{figure}[!ht]
\begin{center}
\includegraphics[width=7.5cm,height=5cm]{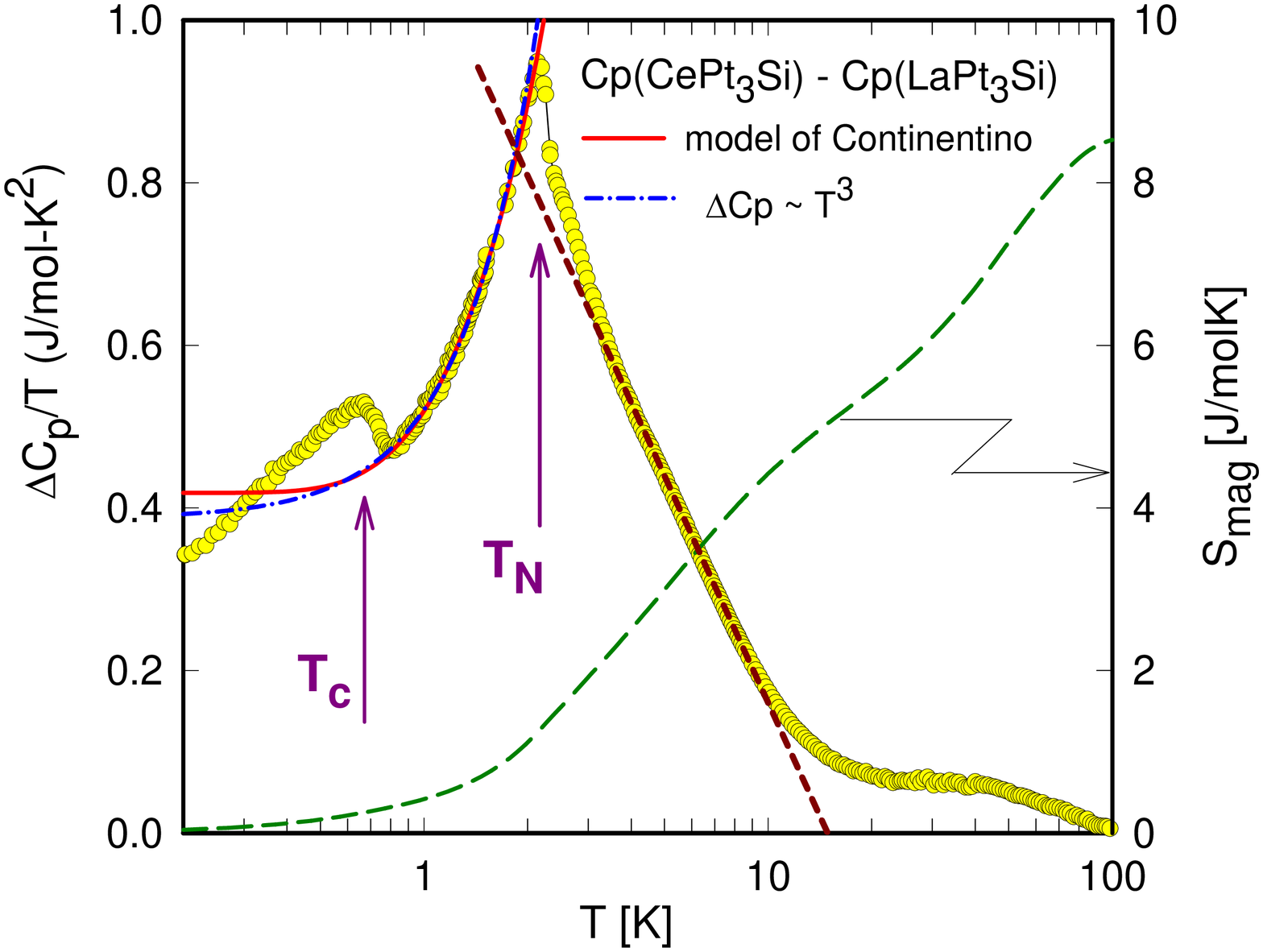}
\end{center}
\caption{Temperature dependent magnetic contribution to the specific
heat, $\Delta C_p$ of $ \rm CePt_3Si$ plotted as $\Delta C_p/T$
on a logarithmic temperature scale. The long-dashed line
represents the magnetic entropy (right axis). The short-dashed
line is a guide to the eyes and roughly 
indicates the non-Fermi liquid behaviour.
The solid line is a fit according to Eq. \ref{eq1} and the
dashed-dotted line is a fit according to $\Delta C_p \sim T^3$.  }
\label{fig1}
\end{figure}

Fig. \ref{fig1} displays $\Delta C_p /T$ vs. $T$
of $\rm CePt_3Si$, where $\Delta C_p$ is defined by the difference
between the $C_p(T)$ data of $\rm CePt_3Si$ and $\rm
LaPt_3Si$, i.e.\ $C_{mag} \sim \Delta C_p$. This plot exhibits three distinct
features: i) the SC transition of $\rm CePt_3Si$ at
$T_c = 0.75$~K (see below), ii) a magnetic transition at $T_N
\approx 2.2$~K and iii) an almost logarithmic tail of $\Delta
C_p/T$ above $T_N$, stretching roughly up to 10~K. Well above 10~K,
Schottky contributions dominate in the specific heat.
The integrated entropy up to 20~K is nearly $R\ln 2$,
 and the entropy of 8.7~J/mol-K integrated 
 up to 100~K is slightly less than $R\ln 4 = 11.5$~J/mol-K.
These results clearly
indicate that the ground state of Ce$^{3+}$ ions is a doublet with
the first excited level above 100 K.
The logarithmic temperature dependence
observed just above the magnetic transition may be considered as
hint of non-Fermi liquid behaviour. Therefore, it is a unique
observation that 
non-Fermi liquid behaviour, magnetic
ordering and eventually a SC transition 
consecutively arises on the same sample upon 
lowering temperature.

In order to analyze in more detail the magnetically ordered region
of the system, a model by Continentino 
\cite{Continentino} is applied with the following analytic expression
for the specific heat well below $T_{mag}$:
\begin{eqnarray}
C_{mag}& = & g \Delta ^{7/2} T^{1/2} \exp (- \Delta / T) \nonumber \\
& \times &
\left[1 +  \frac{39}{20} \left(\frac{T}{\Delta} \right)
+ \frac{51}{32} \left(\frac{T}{\Delta} \right)^2
\right],
\label{eq1}
\end{eqnarray}
This expression is based on  antiferromagnetic magnons with a
dispersion relation given by $\omega = \sqrt{\Delta ^2 + D^2
k^2}$, where $\Delta$ is the spin-wave gap and $D$ is the
spin-wave velocity; $g \propto 1/D^3 \propto 1/\Gamma^3$ and
$\Gamma$ is an effective magnetic coupling between Ce ions. A
least squares fit of Eq. \ref{eq1} to the data below $T_N$ (solid
line, Fig. \ref{fig1}) reveals $\Delta \approx 2.7$~K, a
reasonable gap value with respect to the ordering temperature. 
Another model calculation with simple antiferromagnetic spin waves 
with $C_{mag} \propto T^3$ gives reasonable
agreement with the data, too. A recent neutron diffraction study
on $\rm CePt_3Si$ reveals antiferromagnetic ordering
below $T_N \approx 2.2$~K with a wave vector $ \bf{ k} = \rm
(0,0,1/2)$, i.e. doubling of the magnetic unit cell along $\bf{
c}$-direction \cite{Metoki}. Using  both models, we 
estimate a Sommerfeld coefficient of 0.41 and 0.39~J/molK$^2$
for the former and latter models, respectively. These figures are
in good agreement with an extrapolation of high field specific
heat data where SC is suppressed by applying
magnetic fields.

Considering Kondo type interactions to be responsible for the
significant renormalisation of electrons in $\rm CePt_3Si$  
as evidenced by the
Sommerfeld value $\gamma$, the magnetic entropy allows to
estimate the  Kondo temperature $T_K$. 
Taking the results derived in Ref. \cite{Desgranges}
yields $T_K \approx 7.2$~K.
A second possible estimate $T_K$ follows from
the competition of the RKKY interaction and the Kondo effect,
which leads to a significant reduction of the specific heat 
jump at $T=T_N$. Following the procedure 
developed in Ref. \cite{Besnus}
gives $T_K \approx 9$~K, in reasonable
agreement with the previous estimate.

For better understanding of the magnetic ground state and expected
localized character of Ce {\it 4f} electrons, 
inelastic neutron scattering experiments at the HET spectrometer
of ISIS, UK, were carried out. In order to accurately and reliably
determine magnetic scattering from the magnetic moments
of Ce, both $\rm CePt_3Si$ and $\rm LaPt_3Si$ were investigated in
powder form under identical conditions. For phonon
subtraction, we used two well-established methods \cite{Adroja},
with almost the same results.
Our main finding is that there are strong and clear
features from 10 to 30 meV. 
\begin{figure}[!ht]
\begin{center}
\includegraphics[width=7.5cm,height=5cm]{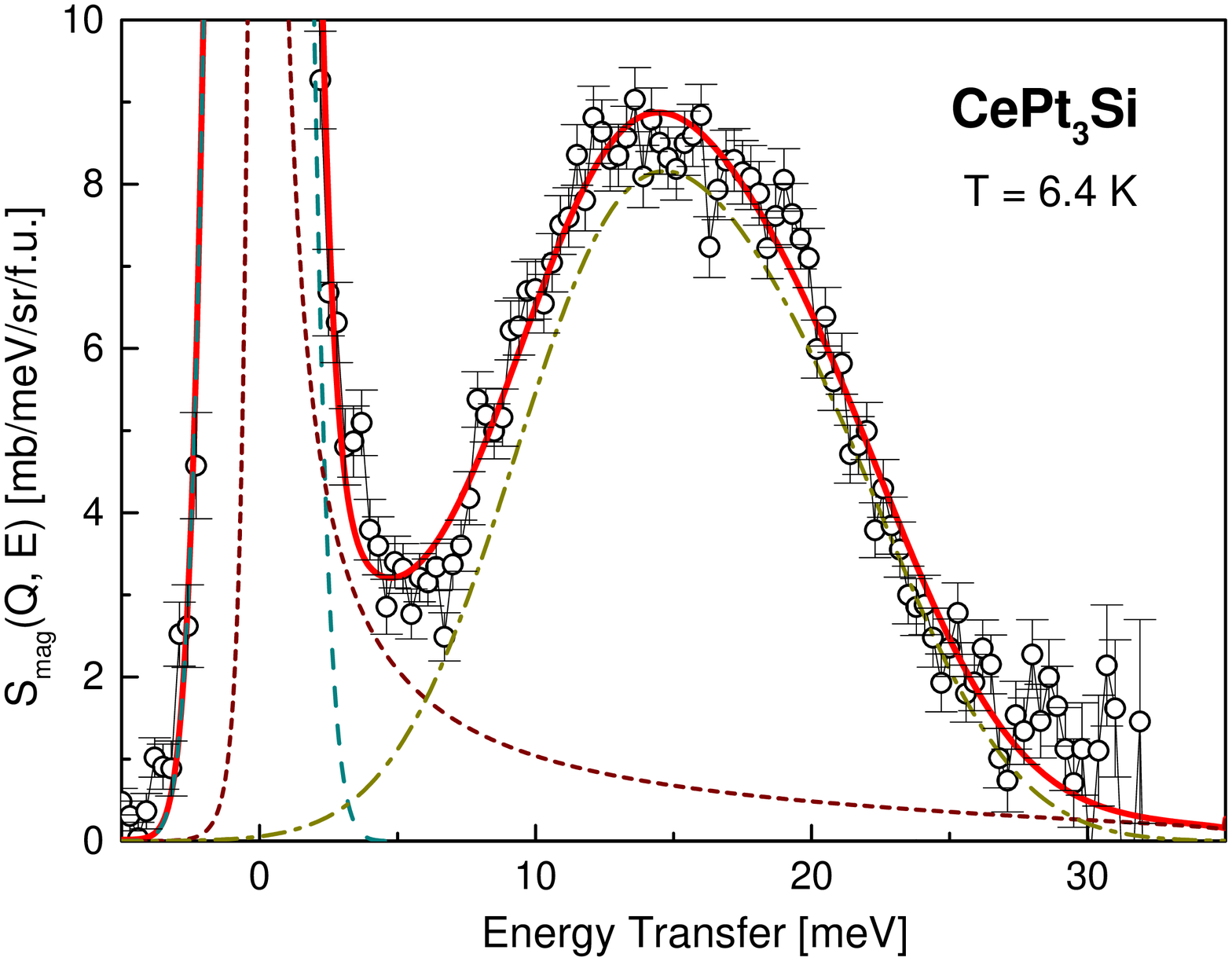}
\end{center}
\caption{Magnetic scattering obtained at 6.4 K with the incident
energy of 35 meV. The dashed line is for the elastic component
with FWHM=2.4 meV while the short-dashed line represents the
quasi-elastic component with FWHM=0.8 meV. The dashed-dotted line is
for the sum of two Lorentzian components centered at 13 and 20 meV
with FWHM=10.0 meV.} 
\label{fig2}
\end{figure}
As shown in Fig. \ref{fig2} 
ought to be of magnetic origin as the phonon contributions have been
subtracted from the data. In order to explain
the data, we used the CEF Hamiltonian for Ce$^{3+}$ with C$_{4v}$
point symmetry: $H_{CEF} = B_2^0 O_2^0 + B_4^0 O_4^0 + B_4^4
O_4^4$. With $B_2^0$=-0.4972 meV, $B_4^0$=0.0418 meV, and
$B_4^4$=0.2314 meV, the observed magnetic
scattering is reasonably well explained
(solid line, Fig. \ref{fig2}). Keeping the CEF
parameters unchanged, we could also explain the data obtained at
94~K equally well. 
Moreover, the two CEF excitations centred at 13 and 20
meV are also consistent with the heat capacity data as discussed
above. We furthermore studied low 
energy excitations using lower incident
energy to find a weak feature around 1.4~meV. 
The dispersion of that intensity at $T = 5$~K, particularly around 
$Q = 0.8$~\AA$^{-1}$ \, is a signature for the 
development of short-ranged magnetic correlations and
explains the anomalous behaviour of the 
specific heat above magnetic ordering. At higher 
temperatures ($T \approx 30$~K) scattering becomes
$Q$-independent. This feature is completely absent in non-magnetic
$\rm LaPt_3Si$. We note that a
recent inelastic neutron scattering experiment \cite{Metoki}
reported two CEF peaks at 1.0 and 24 meV. However, our data show
that their 2nd excitation is most likely to be mistaken.

\subsection*{Superconducting properties of $\rm CePt_3Si$}

\begin{figure}[!ht]
\begin{center}
\includegraphics[width=7.5cm,height=5cm]{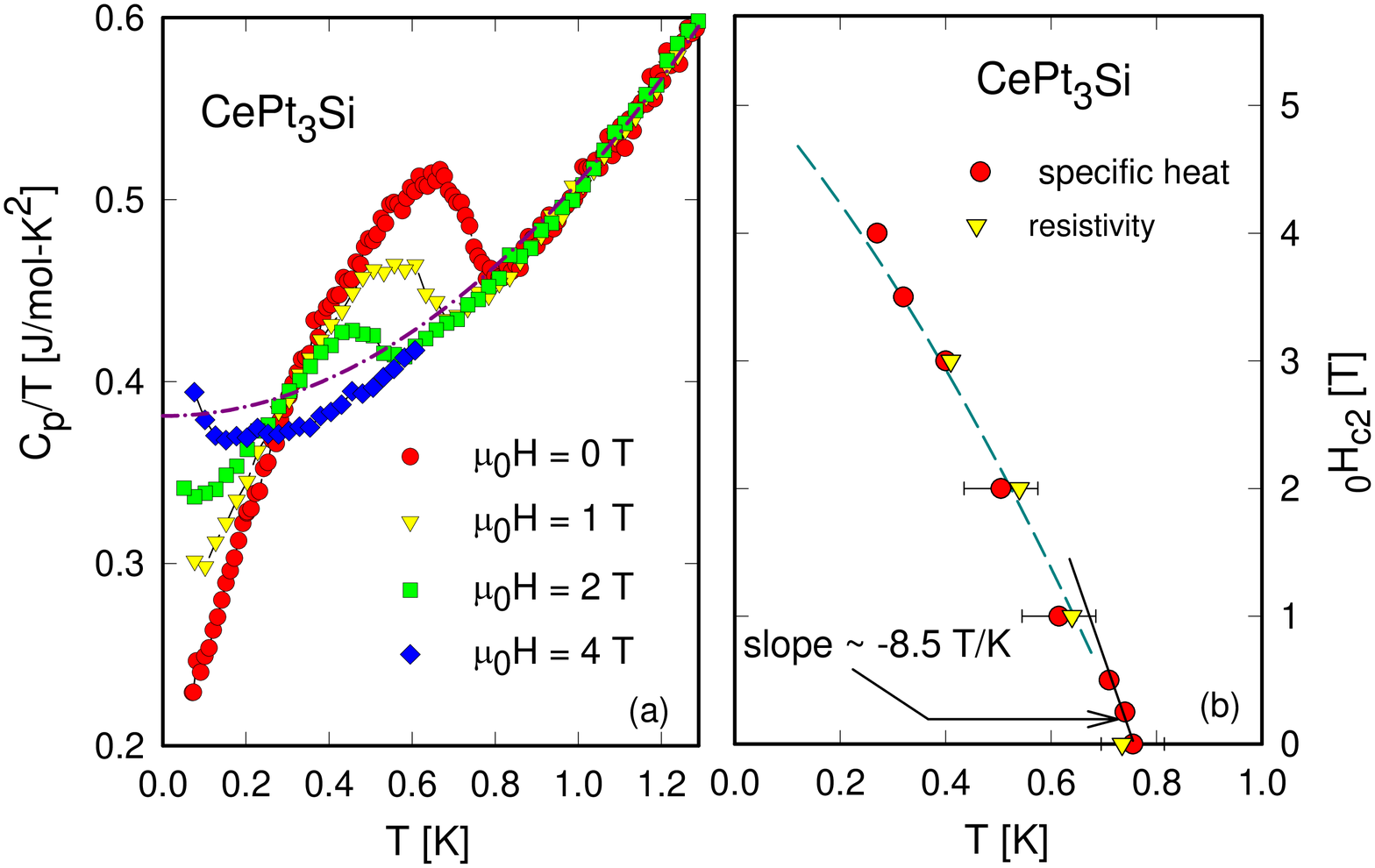}
\end{center}
\caption{(a): Temperature  dependent specific heat $C_p/T$
of $\rm CePt_3Si$ for various values 
of applied fields; the dashed line is a $T^3$ extrapolation
of $C_p(T)$ at 0~T. 
(b): Temperature dependence of
the upper critical field $H_{c2}$. The solid straight line yields 
$H_{c2}' \approx - 8.5$T/K; the dashed line is a guide to the
eyes.  }  
\label{fig3}
\end{figure}

Signs of bulk SC of $\rm CePt_3Si$ below $T_c =
0.75$~K are numerous: zero resistivity, diamagnetic signal in the
susceptibility, a jump in the specific heat and NMR
relaxation rate at $T_c$. Here, we show in Fig. \ref{fig3} (a) the field
dependent specific heat at low temperatures and at various
external magnetic fields.
Sommerfeld coefficient $\gamma_n \approx 0.39$~J/molK$^2$ of $\rm
CePt_3Si$ at zero field can be obtained from a careful
extrapolation of the normal state invoking the $T^3$-dependence
associated with antiferromagnetic ordering. This extrapolation
also satisfies the basic requirement of an entropy balance between
the SC  and normal state regions.

The application of magnetic fields reduces $T_c$, giving rise to a
rather large change of $d H_{c2} /dT \equiv H_{c2}' \approx -
8.5$~T/K, 
in good agreement with the conclusion
drawn from electrical resistivity [see Fig. \ref{fig3}(b)]. An extrapolation
of $T_c (H)$ towards zero  yields $H_{c2}(0) \approx 5$~T,
well above the Pauli - Clogston limiting field \cite{Bauer}.
Furthermore, an estimation of the Sommerfeld coefficient from the
high field data gives 0.36~J/molK$^2$, in fair agreement with
the value obtained from an extrapolation of the normal state in
the zero field data (see Fig. \ref{fig3}). The upturn of $C_p/T$
at lowest temperatures that gets stronger with increasing magnetic
fields is most likely due to the nuclear contribution of
$\rm ^{195}Pt$.

The jump in the specific heat $\Delta C_p/T \vert _{T_c} \approx
0.1$J/molK$^2$, leads to  $\Delta C_p/(\gamma_n
T_c) \approx 0.25$, which is much smaller than 
expected from 
the BCS theory 
($\Delta C_p/(\gamma T_c) \approx 1.43$). Even using
the electronic specific heat coefficient in the SC
state, $\gamma_s \approx 0.18(1)$~J/molK$^2$, we obtained $\Delta
C_p/(\gamma_s T_c) \approx 0.55$ that is still below the BCS
value.

We can think of two scenarios that may well explain such a
substantial reduction of $\Delta C_p/\gamma T_c$ with respect to
the BCS value; i) spin triplet SC like  $\rm
Sr_2RuO_4$ exhibit a similarily reduced magnitude of $\Delta
C_p/(\gamma T_c)$  \cite{Makenzie} and ii) not all electrons
condense into Cooper pairs; 
thus only a fraction of the
carriers mediate the super-current. It implies that
electrons responsible for 
normal state features, like
antiferromagnetic order, coexist with those forming the Cooper
pairs. In fact, the finite value of $\gamma_s \approx
0.18$~J/molK$^2$ provides 
evidence that even
at $T=0$~K a significant portion of the Fermi surface is still not
involved in the SC condensate.

\begin{figure}[!ht]
\begin{center}
\includegraphics[width=7.5cm,height=5cm]{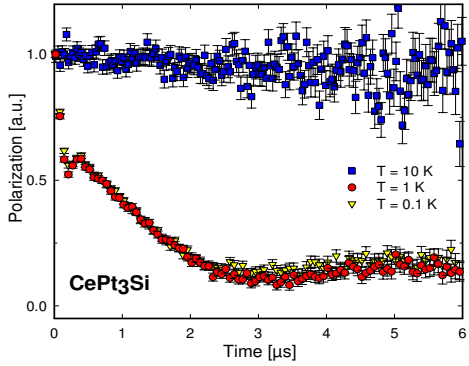}
\end{center}
\caption{Zero field depolarisation rate of $\rm CePt_3Si$
at various temperatures. }
\label{fig4}
\end{figure}

Microscopic evidence for the latter conclusion 
can be found from zero-field $\mu$SR spectroscopy 
data obtained in the magnetic phase below and 
above $T_c$ in the magnetic phase.
(Fig. \ref{fig4}).
At temperatures much above $T_N$, the $\mu$SR signal 
is characteristic of a paramagnetic state with a 
depolarization solely arising from nuclear moments. 
Below $T_N$ the $\mu$SR signal indicates that the 
full sample volume orders magnetically. High statistic 
runs performed above and below $T_c$ did not show
any change of the magnetic signal, supporting the view of a 
microscopic coexistence between magnetism and SC. 
This points to a novel state for SC 
Ce-based heavy-fermion systems at ambient pressure, 
for which, to date, magnetism was found 
to be either absent \cite{petrovic} or strongly 
competing against SC \cite{luke}. 
The observed coexistence is reminiscent of the situation 
observed in UPd$_2$Al$_3$ \cite{feyerherm}, 
where a model of two independent electron subsets, 
localized or itinerant, was proposed 
in view of similar microscopic data \cite{caspary}.

\begin{figure}[!ht]
\begin{center}
\includegraphics[width=7.5cm,height=5cm]{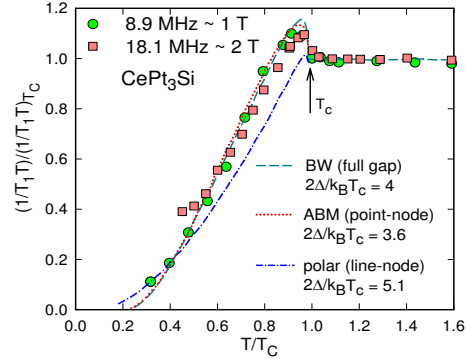}
\end{center}
\caption{A plot of $(1/T_{1}T)/(1/T_{1}T)_{T_c}$ vs. $T/T_{\rm c}$
at 8.9 MHz ($H\sim 1$ T) and 18.1 MHz ($H\sim 2$ T).
 The dashed line is for the
Balian-Werthamer model (BW isotropic triplet SC state)
with a value of $2\Delta/k_{\rm B}T_{\rm c}=4$. 
The dotted line assumes a point-node model with 
$2\Delta/k_{\rm B}T_{\rm c}= 3.6$ and the
dashed-dotted line represents a fit by a line-node gap model
with $2\Delta/k_{\rm B}T_{\rm c}= 5.1$. }
\label{fig5}
\end{figure}

Another microscopic information about the SC state 
can be obtained from the temperature dependent
$^{195}$Pt nuclear spin-relaxation rate $1/T_1$
\cite{Yogi}. Results
are shown as $(1/T_{1}T)_{\rm SC}/(1/T_{1}T)_{\rm N}$ vs.
$T/T_{\rm c}$ plot in Fig. \ref{fig5} for 8.9 and 18.1~MHz. The
relaxation behaviour $1/T_1T$ of $\rm CePt_3Si$ is characterized
by a kind of the Hebel-Slichter anomaly 
\cite{Hebel} indicating coherence
effects as in conventional BCS SC. The peak height,
however, is significantly smaller than that observed for
conventional BCS SC and, additionally, shows no field
dependence at the  8.9 MHz ($H \sim 1$ T) and 18.1 MHz 
($H\sim 2$ T) run.

 $(1/T_{1}T)$ at $H \sim 2$ T seems to  saturate at low
 temperature, which can be
attributed to  the presence of vortex cores where the normal-state
region is introduced. $1/T_{1}T$ at 8.9 MHz ($H \sim 1$ T),
however,  continues to decrease down to $T=0.2$ K, the lowest
measured temperature. Neither an exponential law nor a  $T^{3}$
behaviour is observed for the data down to $T$= 0.2 K. Therefore,
$\rm CePt_3Si$ is the first HF SC that exhibits a peak
in $1/T_1T$ just below $T_{\rm c}$ and, moreover, does not follow
the $T^3$ law reported for most of the unconventional HF
SC (see e.g., Ref. \cite{Tou} and Refs. therein). 

To account for the  relaxation behavior below $T_{\rm c}$ in
 noncentrosymmetric CePt$_3$Si,
three models were adopted for a description of the temperature
dependence of $1/T_1$ at $H\sim 1$ T. The dashed line in Fig.
\ref{fig5} represents a fit according to the  the Balian-Werthamer
model (isotropic spin-triplet SC state) with a value of
$2\Delta/k_{\rm B}T_{\rm c}=3.9$ \cite{Balian}, while the
dashed-dotted line is a fit using a line-node model with $2\Delta/k_{\rm
B}T_{\rm c}= 5.1$. The dotted line refers to a point-node model
with $2\Delta/k_{\rm B}T_{\rm c}= 3.6$.
The  models used, however, failed to give
satisfactory description of the observed temperature dependence of
$1/T_1$ over the entire temperature range. 
While the line-node model gives a reasonable agreement with the
data at lowest temperatures, the BW model describes reasonably well
the data just below $T_{\rm c}$. The peak in $1/T_1T$
would indicate the presence of an isotropic energy gap, even
though a coherence effect - inherent for the isotropic
spin-singlet s-wave pairing state - is absent.

In almost all previous studies on 
either conventional and unconventional SC, 
it was assumed that the crystal has an inversion center, which allows
separate consideration of the even (spin-singlet) and odd
(spin-triplet) components of the SC order parameter.
In CePt$_{3}$Si, however, a center of symmetry  is absent. Therefore, 
the novel relaxation behaviour found below $T_{\rm c}$ hints at
a possibly new class of a SC state being
realized in noncentrosymmetric CePt$_{3}$Si.

Gor'kov and Rashba \cite{Gorkov} demonstrated that in the absence
of inversion symmetry the order parameter becomes a mixture of
spin-singlet and spin-triplet components, which leads, for
instance, to the Knight shift attaining a non- zero value at $T =
0$~K. 
A novel idea with respect to the order parameter of systems
without inversion symmetry was put forward very recently
by Saxena and Monthoux \cite{Montu}.
In their model for the case of broken inversion symmetry, the spins might
rotate in the momentum space around the surface.

\subsection*{Evolution of magnetism and superconductivity in $\rm CePt_3(Si,Ge)$}

In order to follow the evolution of physical properties upon 
Si/Ge substitution, we prepared several alloys. Guinier X-ray
powder intensity profiles of the alloys from the series $\rm
CePt_3(Si_{1-x}Ge_x)$ with $x = 0.02, 0.03, 0.06, 0.10, 0.15$ and
$0.20$ were all indexed completely on the basis of a primitive
tetragonal unit cell, confirming isotypism with the structure type
of $\rm CePt_3Si$ \cite{Sologub}
with a random substitution of the Si/Ge atoms in
the $1a$ sites of space group $P4/mmm$. Small secondary peaks
arise beyond $x \approx 0.2$ indicating the limit of the 
$\rm CePt_3B$ type phase region.
\begin{figure}[!ht]
\begin{center}
\includegraphics[width=7.5cm,height=5cm]{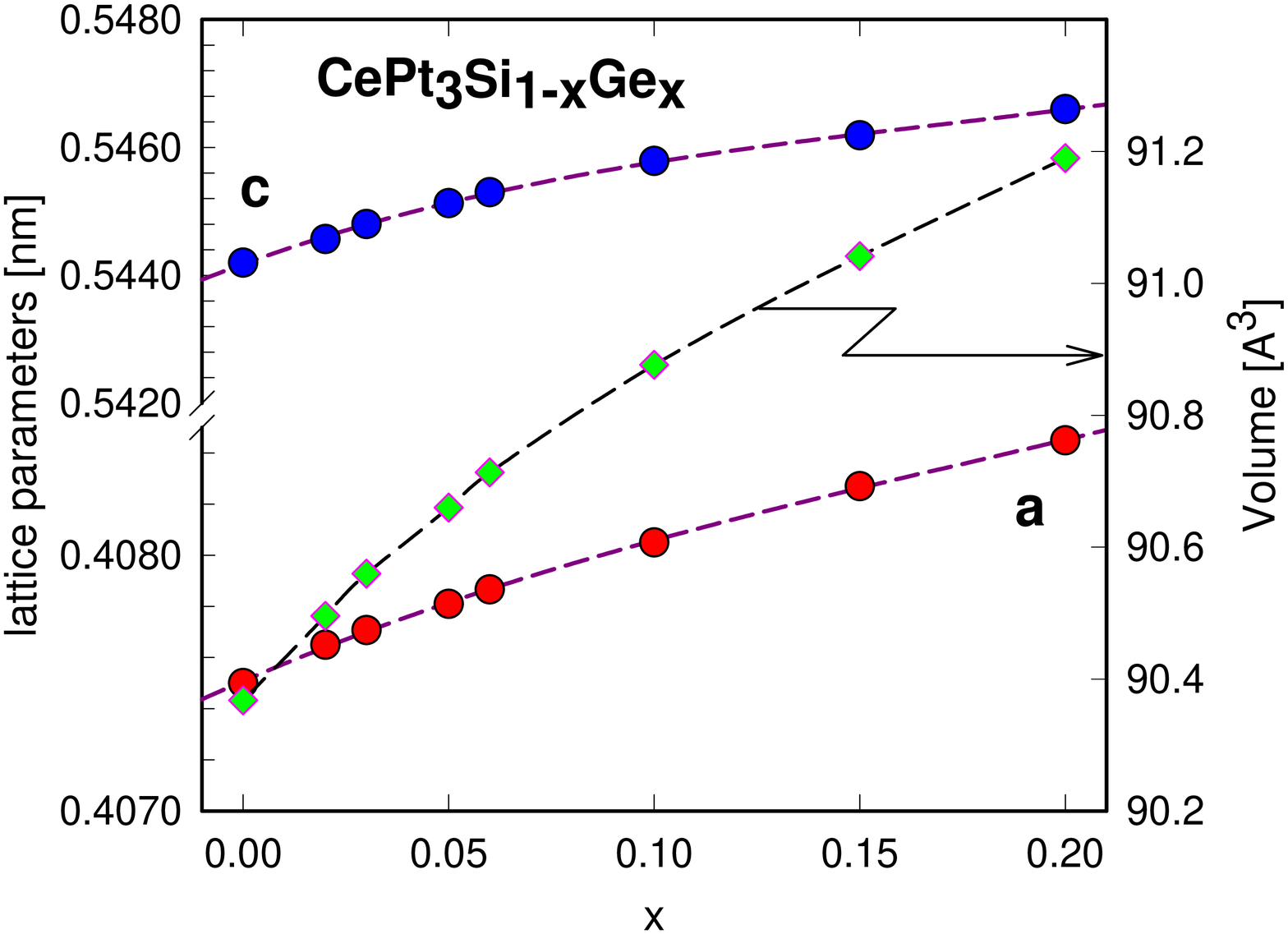}
\end{center}
\caption{Concentration dependent lattice parameters $a$, $c$
and unit cell volume $V$ of $\rm CePt_3Si_{1-x}Ge_x$. }
\label{fig6}
\end{figure}
Figure \ref{fig6} shows the concentration dependent lattice
parameters $a$ and $c$ as a function of Si/Ge substitution,
together with the unit cell volume.  The monotonic increase of
both $a$ and $c$ parameters yields a growing unit cell volume
upon doping. The
general increase of the unit cell volumes reduces chemical
pressure on the Ce atoms leading to a decrease of hybridization.

\begin{figure}[!ht]
\begin{center}
\includegraphics[width=7.5cm,height=5cm]{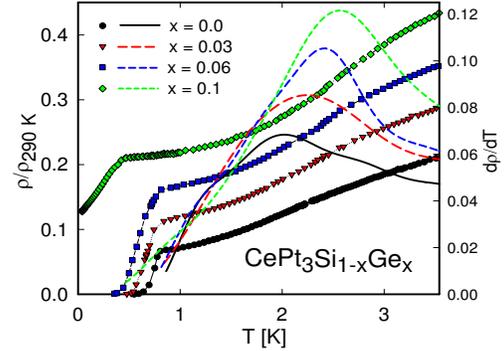}
\end{center}
\caption{Low temperature electrical resistivity $\rho$
for various concentrations $x$ of $\rm CePt_3Si_{1-x}Ge_x$
(symbols), and temperature derivative
d$\rho$/d$T$ (lines, right axis).}
\label{fig7}
\end{figure}
The temperature dependent electrical resistivity  $\rho$ of $\rm
CePt_3(Si_{1-x}Ge_x)$ is shown for several concentrations $x$ in
Fig. \ref{fig7} for  temperatures below about 4~K. The overall
$\rho(T)$ values increase due to increasing substitutional 
disorder on the $1a$ site.
Samples investigated are characterized by a concentration
dependent onset of SC, where the Si/Ge substitution
suppresses the SC roughly above $x = 0.1$. 
At somewhat elevated temperatures, $\rho(T)$ exhibits a pronounced
curvature which can be taken as a signature for long range magnetic
order, in agreement with specific heat measurements. A magnetic
instability is evidenced from the resistivity data in terms of a
d$\rho$/d$T$ plot (lines, Fig. \ref{fig7}, right axis).
d$\rho$/d$T$ shows pronounced anomalies, referring to broadened
magnetic phase transitions; the respective temperatures increase
continuously with increasing Ge content.

\begin{figure}[!ht]
\begin{center}
\includegraphics[width=7.5cm,height=5cm]{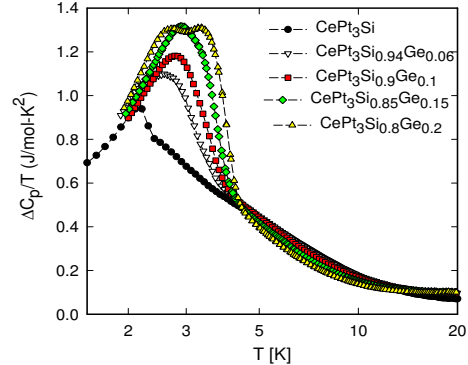}
\end{center}
\caption{Magnetic contribution to the specifc heat $\Delta C_p$
plotted as $C_p/T$ vs. $T$ for various alloys of 
$\rm CePt_3Si_{1-x}Ge_x$. }
\label{fig8}
\end{figure}

The magnetic contribution to the specific
heat of $\rm CePt_3(Si_{1-x}Ge_x)$ is plotted in Fig.
\ref{fig8} as $C_{mag}/T$ vs. $T$ for several compositions up to
$x = 0.2$ from 1.8  to 20 K. $C_{mag}$ is derived as in Fig. 1
using the measured data of isostructural non-magnetic $ \rm
LaPt_3Si$.
A common feature in all the data-sets is a mean-field like anomaly,
which is associated with magnetic order of the cerium sublattice.
$T_N$ rises with
increasing Ge content and, simultaneously, the phase transitions appear to be
broadened, in agreement with the observation made from
d$\rho$/d$T$. While for $x=0$, the logarithmic behaviour well
above $T_N$ is quite pronounced and obvious, 
its temperature range becomes narrower
with increasing Ge content as a consequence of increasing magnetic
interactions, driving a magnetic instability at much higher
temperatures. The double-peak feature 
for $x = 0.2$ is most likely due to a secondary phase as
indicated by our XRD analysis.

\begin{figure}[!ht]
\begin{center}
\includegraphics[width=7.5cm,height=5cm]{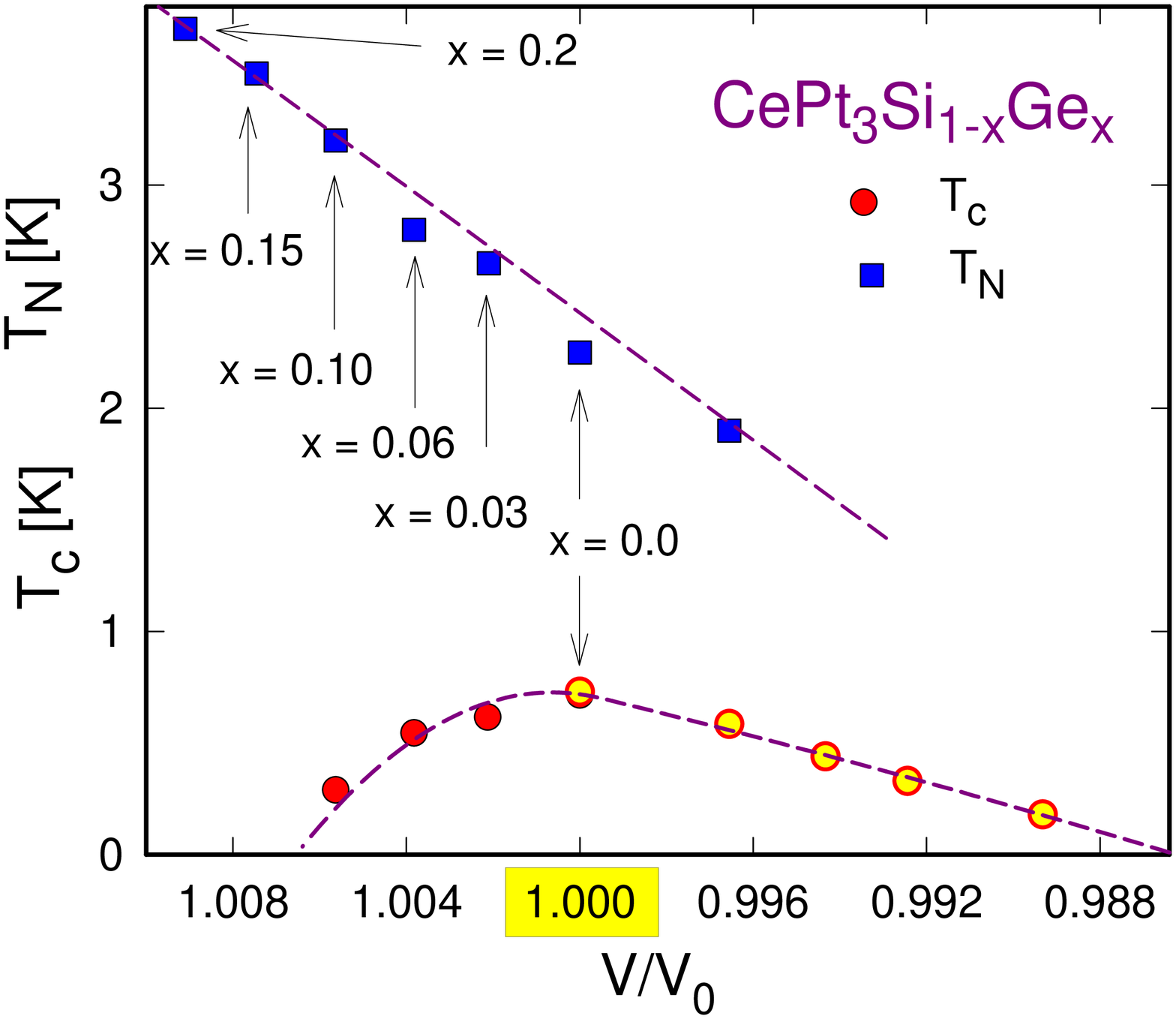}
\end{center}
\caption{Phase diagram of $\rm CePt_3(Si_{1-x}Ge_x)$. }
\label{fig9}
\end{figure}

The phase diagram shown in Fig. \ref{fig9} summarizes
characteristic temperatures deduced for $\rm CePt_3(Si_{1-x}Ge_x)$
at ambient pressure, as well as data derived from resistivity
studies on $\rm CePt_3Si$ under hydrostatic pressure up to
about 15~kbar \cite{Nicklas}. In order to make comparison between
substitution and pressure, Murnaghan's equation of state is
adopted, with a bulk modulus $B_0 = 1000$~kbar.
Pressure of 15~kbarcorresponds then with a reduction of the unit
cell volume of about 1~\%.
The phase  diagram thus deduced resembles very well those
characteristics, which determine the standard generic phase
diagram associated with a quantum phase transition. The increase
of the unit cell volume by the Si/Ge substitution is, in a
first-order approximation, responsible for a decrease of
hybridisation. This has two consequences: i) the Kondo interaction
decreases and ii) magnetic interaction strengthens, causing the
observed increase of the magnetic transition temperature.
Concomitantly, the SC transition temperature becomes
suppressed and finally, SC vanishes beyond $x > 0.1$. Increasing
hydrostatic pressure is responsible for a decrease of both $T_N$
and $T_c$. Ternary $\rm CePt_3Si$ thus appears to
be, by chance, situated at the maximum position of the
``superconducting dome''. Whether or not this SC dome is
constrained within the magnetic phase is still unknown. Depending
on the particular choice of $B_0$, i.e. smaller or larger
than $B_0 = 1000$~kbar the volume below $V/V_0 =
1$ becomes strechted or narrower. In any case, $T_c^{max}$ is well
below the magnetic phase line, being a signature that Cooper
pairing may be mediated by magnetic fluctuations rather than by
the standard phonon mechanism.

We summarize that non-centrosymmetric $\rm CePt_3Si$ is a heavy
fermion SC with $T_c = 0.75$~K that orders
magnetically at $T_N = 2.2$~K. The NMR relaxation rate $1/T_1$
shows unexpected features which were found before neither in
conventional nor in heavy fermion SC, indicative of very unusual
shapes of the SC order parameter. In fact, a number
of theoretical scenarios support these observations. The Si/Ge
substitution simply drives a  volume expansion,
thus magnetism is stabilized and SC finally ceases to
exist.

Work supported by the Austrian FWF P16370, P15066
and by the DFG, SFB 484 (Augsburg).  
The Austrian - Russian exchange within project I.18/4
is acknowledged. JGP and WYS acknowledge
financial support by the CSCMR of Seoul National University.

\end{document}